# Low-threshold optically pumped lasing in highly strained Ge nanowires

Shuyu Bao[1,2†], Daeik Kim[3†], Chibuzo Onwukaeme[3†], Shashank Gupta[4†], Krishna Saraswat[4], Kwang Hong Lee[2], Yeji Kim[3], Dabin Min[3], Yongduck Jung[3], Haodong Qiu[1], Hong Wang[1], Eugene A. Fitzgerald[2], Chuan Seng Tan[1,2*] and Donguk Nam[1,3*]

[1]School of Electrical and Electronic Engineering, Nanyang Technological University, 50 Nanyang Avenue, Singapore 639798

[2] Singapore-MIT Alliance for Research and Technology (SMART), 1 CREATE Way, #09-01/02 CREATE Tower, Singapore 138602

[3]Department of Electronic Engineering, Inha University, Incheon 402-751, South Korea

[4]Department of Electrical Engineering, Stanford University, Stanford, California 94305, USA

[†]These authors contributed equally to this work

[*]E-mail: tancs@ntu.edu.sg; dnam@ntu.edu.sg

The integration of efficient, miniaturized group IV lasers into CMOS architecture holds the key to the realization of fully functional photonic-integrated circuits[1,2]. Despite several years of progress, however, all group IV lasers reported to date exhibit impractically high thresholds owing to their unfavorable bandstructures[3–8]. Highly strained germanium with its fundamentally altered bandstructure has emerged as a potential low-threshold gain medium[9–24], but there has yet to be any successful demonstration of lasing from this seemingly promising material system. Here, we demonstrate a low-threshold, compact group IV laser **that employs germanium nanowire under a 1.6% uniaxial tensile strain as the gain medium. The amplified material gain in strained germanium can sufficiently surmount optical losses at 83 K, thus allowing the first observation of multimode lasing with an optical pumping threshold density of ~3.0 kW cm$^{-2}$. Our demonstration opens up a new horizon of group IV lasers for photonic-integrated circuits.**

Approximately a decade ago, germanium (Ge) began attracting much attention as a potential gain medium for integrated group IV lasers, which still remain as the last missing piece for fully functional photonic-integrated circuits (PICs)[25]. Unlike silicon (Si), the energy difference between the direct Γ and indirect L conduction valleys of Ge is only ~140 meV, and therefore it is possible to force some electrons to populate the direct Γ conduction valley and to create a population inversion in Ge[25]. By employing heavy n-type doping in ~0.2% biaxially strained Ge to assist with population inversion, optically and electrically pumped Ge-on-Si lasers have been demonstrated[3–5]. However, these lasers suffered from an extremely high lasing threshold current density (~280 kA cm$^{-2}$) that precludes practical integration. It was well known that intrinsic losses from free carriers were a major culprit for this inefficiency[25,26], which other researchers identified more specifically the inter-valence band absorption (IVBA) process from free holes[27,28].

During the last few years, researchers have been making relentless efforts to overcome the drawback of indirect bandgap Ge through a variety of techniques. Among those, tin (Sn) alloying into Ge has shown great promise because GeSn alloy with a reasonably high Sn concentration (>6.5%) can achieve a direct bandgap[6–8,29–31]. Recently, a Fabry-Perot type GeSn waveguide with 12.6% Sn demonstrated lasing with an optical pumping threshold density of 325 kW cm$^{-2}$ at 20 K[6], and a couple of more successful demonstrations have since followed[7,8]. However, the lasing threshold of those GeSn lasers was still too high to be practical possibly owing to faster than desired non-radiative recombination processes[6].

On the other hand, the application of large mechanical tensile strain can also address the challenge of the indirect bandgap in Ge by fundamentally altering the bandstructure[10–16]; the small energy difference between the Γ and L conduction valleys can be reduced even further with tensile strain, resulting in an increased material gain[15,16]. This hypothesis has been supported by numerous theoretical simulations that predict a significant reduction in lasing threshold through tensile strain[26,32]. With regard to practical realization, several innovative platforms for inducing large mechanical strain have been experimentally demonstrated[10–14], thus suggesting the possibility of low-threshold group IV lasers.

While lasing from highly strained Ge has been pursued intensively in several recent reports[17–24], there has been no successful demonstration of optical amplification in such a seemingly promising material system. Most notably, El Kurdi et al. investigated photoluminescence from direct bandgap Ge under a 1.75% biaxial tensile strain; however, no lasing action was observed even at cryogenic temperature, possibly owing to a non-uniform strain distribution in the gain medium and/or a poor thermal conduction arising from suspended structures[21,23]. The authors also performed theoretical calculations on the actual device temperature under the same pump condition used in their measurements[23], and found that the temperature rises to 210 K which is significant enough to preclude lasing because of the increased material loss. This temperature dependent optical characteristic is discussed theoretically and experimentally throughout this study.

In this Letter, we present the first observation of low-threshold lasing from Ge nanowires under 1.6% uniaxial tensile strain. The unique design of our Ge laser possesses several advantages such as a highly homogeneous strain distribution and an excellent optical mode overlap with the gain medium. In contrast to the conventional geometry in which the strained Ge gain medium is suspended in air[19–21,23], our laser structures hold the strained Ge nanowire in close contact with silicon dioxide ($SiO_2$), which provides superior thermal conduction to the air gap while confining optical fields within the Ge layer (Supplementary Section 1). Figure S1d illustrates the comparison between the thermal properties of the suspended and stiction structures. While the temperature of the suspended structure quickly increases to 180 K at ~ 7 kW cm$^{-2}$ similarly to ref 23, our laser structure shows an almost negligible temperature increase with the same pumping condition. With a pulsed optical pumping threshold density of ~3.0 kW cm$^{-2}$, we observe an unambiguous multimode lasing action, which is evidenced by clear threshold behaviors in output power and linewidth as a function of pump power. Theoretical modeling shows that the IVBA, the recently discovered dominant loss factor in Ge[27,28], is markedly reduced at cryogenic temperature[33] and therefore can be compensated by the amplified material gain in strained Ge, leading to the development of a large optical net gain of ~415 cm$^{-1}$. Temperature-dependent measurements (Fig. 4) also confirm the critical role of thermal management in suppressing the material loss as demonstrated in our laser devices, which is in contrast to the earlier work[23] showing no evidence of lasing from suspended structures with a poor thermal management.

The laser structure consists of a Ge nanowire gain medium under 1.6% uniaxial tensile strain along the <100> direction, which is surrounded by two stressing pads containing distributed Bragg reflectors (DBRs) (Fig. 1a). The strain amplification method first introduced in ref [34] allows for a convenient tuning of the strain level by changing the length of the stressing pads (Supplementary Section 2). The nanowire is optically pumped by a 1064-nm pulsed laser, which allows the emitted photons to strongly oscillate between a pair of DBRs (see Methods). The bottom inset of Fig. 1a shows the cross-sectional transmission electron microscopy (TEM) image of the Ge-on-insulator (GOI) substrate used in the present research (Supplementary Section 3). Figure 1b illustrates the strong spatial overlap between optical and strain fields in our design. The strain map measured by Raman spectroscopy (middle) shows a highly uniform strain distribution of ~1.6% uniaxial tension over the entire nanowire, which in turn allows a homogeneous gain property (see Methods). The contour of an SEM image (top) is superimposed onto the strain map for clarity. The calculated optical field (bottom) is strongly confined to the nanowire gain medium, thus enabling a large optical confinement factor of ~0.45 according to finite-difference time-domain (FDTD) simulation. Figure 1c presents the calculated bandstructure of Ge under 1.6% uniaxial tensile strain[28]. Two major optical processes, gain and IVBA, are labeled to highlight that gain should overcome IVBA for the occurrence of lasing. The large mechanical strain not only reduces the energy difference between the $\Gamma$ and L conduction valleys but also lifts the degeneracy of the top two valence bands (VB1 and VB2).

We successfully observe multimode lasing by optically exciting a 1.6%-strained Ge nanowire integrated with DBRs (Fig. 1a) at a low temperature of 83 K. Figure 2a presents the emission spectra from the laser structure at different pump powers. A linear polarizer inclined perpendicular to the nanowire axis is placed in front of the spectrometer to investigate polarization-dependent lasing behavior.

At a low pump power of 0.7 kW cm$^{-2}$, the emission shows broad spontaneous emission which consists of two overlapped peaks arising from optical transitions of Γ-VB1 and Γ-VB2[22,35]. At an increased pump power of 1.4 kW cm$^{-2}$, visible longitudinal cavity modes owing to the presence of DBRs emerge on the long wavelength side as the optical loss is gradually compensated by the material gain. Although such cavity modes near the band edge from highly strained Ge have been already observed numerous times[18–24], all the previous reports showed the apparent intensity saturation and linewidth broadening of cavity modes with pump powers possibly owing to insufficient gain at low strain[20,24], poor optical cavity[21,22] and/or unsatisfactory thermal management[18,23].

In the present study, on the other hand, the cavity modes near the band edge attain a superlinear increase in intensity along with a significant linewidth reduction as the pump power is increased to 3.5 kW cm$^{-2}$. This phenomenon clearly presents the evidence of optical amplification in highly strained Ge gain medium. Interestingly, the spontaneous emission well above the band edge (<1550 nm) also collapses into sharp cavity modes, thus suggesting that the entire emission band is close to the transparency condition. The full-width at half-maximum (FWHM) of the emission band at this pump power is ~150 nm (marked as two blue arrows). The mode spacing is measured to be ~8 nm, which is consistent with FDTD simulation (Supplementary Section 4). It is worth mentioning that the occurrence of amplified cavity modes over the whole emission band is in stark contrast to conventional lasing materials for which cavity modes are generally observed only within a relatively narrow gain bandwidth. This unique behavior in highly strained Ge will be explained in the following gain modeling section.

At higher pump powers of 5.1, 8.5, and 14.6 kW cm$^{-2}$, only the modes near 1530 nm continue growing rapidly in intensity as they enter the lasing regime while the modes near the band edge (>1600 nm) appear to saturate. At the highest pump power of 14.6 kW cm$^{-2}$, the cavity modes near 1530 nm obtain >20x higher intensity than the background spontaneous emission, and exhibit a well-defined Gaussian shape following the net gain spectrum with the FWHM bandwidth of <50 nm while the bandwidth of the background emission is ~150 nm. The large bandwidths of both spontaneous emission[22,23,35] and net gain[36] have been previously observed in highly strained Ge, and can be attributed to the unusually large amount of strain-induced valence band splitting (~72 meV for 1.6% uniaxial strain). The FWHM of a single cavity mode at 1530 nm is ~1.3 nm, corresponding to a Q-factor of >1,100.

Lasing action can also be quantitatively evidenced by the nonlinear behavior of the integrated output intensity within the gain bandwidth as a function of pump power (Fig. 2b). The output intensity shows a superlinear growth near the threshold pump power (~3.0 kW cm$^{-2}$) followed by a linear growth in the lasing regime, which clearly indicates a distinctive lasing action. The corresponding double-logarithmic plot of the output intensity dependence on the pump power (Fig. 2b, inset) manifests a typical nonlinear threshold behavior. Figure 2c illustrates the linewidth evolution of the highest lasing mode at 1530 nm as a function of pump power. The linewidth reduces from ~2.5 nm to

~1.3 nm above the threshold pump power. The linewidth and the corresponding Q-factor at the threshold pump power of ~3.0 kW cm$^{-2}$ is ~1.8 nm and ~850, respectively, assuming the effective index of 3.2 which is calculated from FDTD simulations. The threshold gain is estimated to be 336 cm$^{-1}$ with the simulated optical confinement factor of 0.45. Above 15 kW cm$^{-2}$, the output intensity starts saturating while the linewidth broadens, which can be attributed to the sample heating. This challenge may be overcome by enhancing the thermal conductivity of the Ge nanowire structure through replacing the underlying insulating layer with a more thermally conductive material such as silicon nitride ($Si_3N_4$) and aluminum oxide ($Al_2O_3$).

The polarization dependence of the laser emission is also investigated as shown in Fig. 2d. The emission polarized parallel to the nanowire axis (blue) does not show the evidence of optical amplification, manifesting a strong anisotropic gain property caused by uniaxial tensile strain. Figure 2e presents the emission spectra from an unstrained Ge nanowire for 0.7 kW cm$^{-2}$ (black) and 14.6 kW cm$^{-2}$ (red). In contrast to the 1.6%-strained nanowire exhibiting obvious lasing behaviors, the spectra from the unstrained structure do not exhibit such features of lasing which can be attributed to the absence of optical net gain achieved in unstrained Ge.

Theoretical modeling is performed for a comprehensive understanding of the gain and loss dynamics in the strained Ge material system. We use the empirical pseudopotential method (EPM) for computing the bandstructure of highly strained Ge, and numerical analysis is employed to calculate the material gain via Fermi's golden rule[28] (Supplementary Section 5). For IVBA, we use experimentally extracted absorption cross-sections for room temperature[11,27] and cryogenic temperature[33]. The loss is the sum of free electron absorption (FEA) and IVBA, and IVBA is the dominant factor as previously discovered in refs 11 and 28 (Supplementary Fig. S5a). At $7 \times 10^{19}$ cm$^{-3}$ carrier density, IVBA is ~214 cm$^{-1}$ whereas FEA is ~96 cm$^{-1}$ for the experimental gain peak wavelength of ~1530 nm, thus making IVBA the major loss term.

Figure 3a presents the calculated material gain with reverse sign (solid lines) and loss (dashed lines) at 83 K for three different carrier injection densities of $4 \times 10^{19}$ cm$^{-3}$ (blue), $5 \times 10^{19}$ cm$^{-3}$ (green) and $8 \times 10^{19}$ cm$^{-3}$ (red).

At an injection density of $4 \times 10^{19}$ cm$^{-3}$, the material gain attains its peak intensity above 1600 nm, which corresponds to the optical transition between the Γ conduction valley and the top valence band, VB1. As the injection density is increased to $5 \times 10^{19}$ cm$^{-3}$, a large number of holes start filling up the second valence band, VB2, resulting in the emergence of another gain peak below 1600 nm. The significantly increased gain bandwidth extends down to ~1500 nm, thus explaining the experimental observation of cavity modes within such a large spectral range for the spectrum taken at 3.5 kW cm$^{-2}$ in Fig. 2a. The gain continues increasing at an injection density of $8 \times 10^{19}$ cm$^{-3}$ and allows the achievement of a large optical net gain of ~415 cm$^{-1}$ at ~1510 nm. The gain for unstrained Ge at an injection density of $8 \times 10^{19}$ cm$^{-3}$ (black solid line) is overwhelmed by the loss for the same injection density (red dashed line), highlighting the significance of tensile strain for obtaining optical net gain in Ge.

Figure 3b shows the net gain spectrum obtained by subtracting the loss from the gain for an injection density of $8 \times 10^{19}$ cm$^{-3}$. Although there exist two gain bands associated with optical transitions of Γ-VB1 and Γ-VB2, the gain band centered at ~1510 nm only surmounts the experimental threshold gain of 336 cm$^{-1}$. The measured gain bandwidth is also presented as a blue arrow, which is in reasonable agreement with the theoretical gain bandwidth. The slight discrepancy between the theoretical and experimental gain bandwidths may be ascribed to the uncertainty in bandstructure calculation for such highly strained Ge.

Figure 4 presents the integrated output intensity of the 1.6%-strained Ge nanowire as a function of the pump power at different temperatures of 83, 123, and 173 K. While at 83 K, optical amplification enables the output intensity to grow superlinearly near the threshold, the datasets for 123 K and 173 K do not show such a clear nonlinear threshold behavior. The inset shows the normalized emission spectra taken at 83 K (red) and 173 K (black) at a pump power of 14.6 kW cm$^{-2}$. In contrast to the spectrum for 83 K showing the dominant lasing modes over the background spontaneous emission, the cavity modes are highly suppressed at 173 K and do not show a superlinear growth in intensity at higher pump powers, which displays a strong temperature dependence of gain characteristics in strained Ge. The gain and loss analysis for 300 K (Supplementary Fig. S5b) shows no optical net gain achievable in 1.6%-strained Ge owing to a substantial IVBA, thus highlighting the significant role of IVBA in realizing Ge lasers.

Recent theoretical studies predicted that room-temperature lasing can be achieved from >3% strained Ge because of the further amplified material gain[11,28]. Since higher than 3% uniaxial tensile strain has been already achieved in Ge nanostructures[11,12], we anticipate that the further development of our laser structure will certainly enable us to obtain such highly strained Ge gain media coupled with high-Q optical cavity for the realization of room-temperature Ge lasers.

In summary, we have demonstrated the first observation of lasing from highly strained Ge nanowires. Our nanowire laser design secures robust mechanical, optical, and thermal properties, which played major roles in enabling the previously unattainable low-threshold lasing from highly strained Ge. Excitation-, temperature-, and polarization-dependent optical characterizations showed the achievement of pulsed lasing at 83 K with an optical pumping threshold density of ~3.0 kW cm$^{-2}$, the lowest value among all group IV lasers reported to date[3–8]. Room temperature lasing could be achieved in our architecture, but with a higher level of tensile strain that is within reach[11,12,28]. We believe that single mode operation can be realized by further optimizing the cavity through the use of shorter optical cavities or employing distributed feedback (DFB) grating. The realization of a low-threshold, electrically pumped Ge laser is also possible by using a lateral p-i-n junction[37]. In addition, the ability to tune the strain conveniently by a conventional lithography[11,14,34] should enable the creation of multiple lasers operating at different spectral ranges on a single chip, which is crucial for wavelength division multiplexed optical interconnects. Our demonstration of a low-threshold, highly strained Ge nanowire laser markedly narrows the gap between conventional III–V lasers and their group IV counterparts, thus opening up new avenues for PICs.

**Methods**

A micro-Raman spectroscopy system with a 532-nm laser source was used to measure the Raman peak shift of a strained Ge laser structure. By using Lorenz fit of the peak shift with a uniaxial strain-shift coefficient of 152 cm$^{-1}$, the strain value can be derived from the Raman measurement. Raman mapping was conducted to measure the strain distribution over the laser structure. The piezo stage step size was set to 200 nm during the mapping.

For photoluminescence spectroscopy, a 1064-nm pulsed laser was focused using a 50x magnification lens. The pulse duration and repetition period were 20 ns and 100 ns (duty cycle of 20%), respectively, to minimize heating. All pump power values refer to the peak laser intensity considering 20% duty cycle. The spot size was set to ~15 μm to uniformly illuminate 8 μm long nanowires. The scattered light from a DBR was collected through the same objective lens, coupled to a spectrometer and detected by an InGaAs 1D-array detector with the cut-off at 1.7 μm.

**References**

1. Soref, R. The past, present, and future of silicon photonics. *IEEE J. Sel. Top. Quantum Electron.* **12,** 1678–1687 (2006).

2. Zhou, Z., Yin, B. & Michel, J. On-chip light sources for silicon photonics. *Light Sci. Appl.* **4,** 1–13 (2015).

3. Liu, J., Sun, X., Camacho-Aguilera, R., Kimerling, L. C. & Michel, J. Ge-on-Si laser operating at room temperature. *Opt. Lett.* **35,** 679–81 (2010).

4. Camacho-Aguilera, R. E. *et al.* An electrically pumped germanium laser. *Opt. Express* **20,** 11316–20 (2012).

5. Koerner, R. *et al.* Electrically pumped lasing from Ge Fabry-Perot resonators on Si. *Opt. Express* **23,** 14815 (2015).

6. Wirths, S. *et al.* Lasing in direct-bandgap GeSn alloy grown on Si. *Nat. Photonics* **9,** 88–92 (2015).

7. Stange, D. *et al.* Optically pumped GeSn microdisk lasers on Si. *ACS Photonics* **3,** 1279–1285 (2016).

8. Al-kabi, S. *et al.* An optically pumped 2.5 μm GeSn laser on Si operating at 110 K. *Appl. Phys. Lett.* **109,** 171105 (2016).

9. Sánchez-Pérez, J. R. *et al.* Direct-bandgap light-emitting germanium in tensilely strained nanomembranes. *Proc. Natl. Acad. Sci. U. S. A.* **108,** 18893–8 (2011).

10. Jain, J. R. *et al.* A micromachining-based technology for enhancing germanium light emission via tensile strain. *Nat. Photonics* **6,** 398–405 (2012).

11. Suess, M. J. *et al.* Analysis of enhanced light emission from highly strained germanium microbridges. *Nat. Photonics* **7,** 466–472 (2013).

12. Sukhdeo, D. S., Nam, D., Kang, J.-H., Brongersma, M. L. & Saraswat, K. C. Direct bandgap germanium-on-silicon inferred from 5.7% <100> uniaxial tensile strain. *Photonics Res.* **2,** A8 (2014).

13. Capellini, G. *et al.* Strain analysis in SiN/Ge microstructures obtained via Si-complementary metal oxide semiconductor compatible approach. *J. Appl. Phys.* **113,** 013513 (2013).

14. Nam, D. *et al.* Strain-induced pseudoheterostructure nanowires confining carriers at room temperature with nanoscale-tunable band profiles. *Nano Lett.* **13,** 3118–3123 (2013).

15. El Kurdi, M., Fishman, G., Sauvage, S. & Boucaud, P. Band structure and optical gain of tensile-strained germanium based on a 30 band k·p formalism. *J. Appl. Phys.* **107,** 013710 (2010).

16. Virgilio, M., Manganelli, C. L., Grosso, G., Pizzi, G. & Capellini, G. Radiative recombination and optical gain spectra in biaxially strained n-type germanium. *Phys. Rev. B* **87,** 235313 (2013).

17. Prost, M. *et al.* Tensile-strained germanium microdisk electroluminescence. *Opt. Express* **23,** 6722 (2015).

18. Petykiewicz, J. *et al.* Direct bandgap light emission from strained germanium nanowires coupled with high‐Q nanophotonic cavities. *Nano Lett.* **16,** 2168–2173 (2016).

19. Geiger, R. Direct band gap germanium for Si-compatible lasing. PhD dissertation, ETH Zürich (2016).

20. Al-Attili, A. Z. *et al.* Whispering gallery mode resonances from Ge micro-disks on suspended beams. *Front. Mater.* **2,** 1–9 (2015).

21. Kurdi, M. El *et al.* Direct band gap germanium microdisks obtained with silicon nitride stressor layers. *ACS Photonics* **3,** 443–448 (2016).

22. Ghrib, A. *et al.* All-around SiN stressor for high and homogeneous tensile strain in germanium microdisk cavities. *Adv. Opt. Mater.* **3,** 353–358 (2015).

23. Kurdi, M. El *et al.* Tensile-strained germanium microdisks with circular Bragg reflectors. *Appl. Phys. Lett.* **108,** 091103 (2016).

24. Xu, X., Hashimoto, H., Sawano, K. & Maruizumi, T. Highly n-doped germanium-on-insulator microdisks with circular Bragg gratings. *Opt. Express* **25,** 1279–1285

(2017).

25. Liu, J. *et al.* Tensile-strained, n-type Ge as a gain medium for monolithic laser integration on Si. *Opt. Express* **15,** 11272–7 (2007).

26. Dutt, B. *et al.* Roadmap to an efficient germanium-on-silicon laser: strain vs. n-type doping. *IEEE Photonics J.* **4,** 2002–2009 (2012).

27. Carroll, L. *et al.* Direct-gap gain and optical absorption in germanium correlated to the density of photoexcited carriers, doping, and strain. *Phys. Rev. Lett.* **109,** 057402 (2012).

28. Gupta, S. *et al.* Dramatic and previously overlooked interaction between strain and parasitic absorption in germanium with major implications for Si-compatible lasing. in *CLEO Sci. Innov.* **SW1M.4** (2016).

29. Costa, V. R. D. *et al.* Optical critical points of thin-film $Ge_{1-y}Sn_y$ alloys: A comparative $Ge_{1-y}Sn_y/Ge_{1-x}Si_x$ study. *Phys. Rev. B* **73,** 125207 (2006).

30. Chen, R. *et al.* Increased photoluminescence of strain-reduced, high-Sn composition $Ge_{1-x}Sn_x$ alloys grown by molecular beam epitaxy. *Appl. Phys. Lett.* **99,** 181125 (2011).

31. Gupta, S., Magyari-köpe, B., Nishi, Y. & Saraswat, K. C. Achieving direct band gap in germanium through integration of Sn alloying and external strain. *J. Appl. Phys.* **113,** 073707 (2013).

32. Peschka, D. *et al.* Modeling of edge-emitting lasers based on tensile strained germanium microstrips. *IEEE Photonics J.* **7,** 1502115 (2015).

33. Morozov, I. & Ukhanov, I. Effect of doping on the absorption spectra of p-Ge. *Sov. Phys. J.* **13,** 744–747 (1970).

34. Minamisawa, R. A. *et al.* Top-down fabricated silicon nanowires under tensile elastic strain up to 4.5%. *Nat. Commun.* **3,** 1096 (2012).

35. Nam, D. *et al.* Study of carrier statistics in uniaxially strained Ge for a low-threshold Ge laser. *IEEE J. Sel. Top. Quantum Electron.* **20,** 1500107 (2014).

36. de Kersauson, M. *et al.* Optical gain in single tensile-strained germanium photonic wire. *Opt. Express* **19,** 17925–34 (2011).

37. Ellis, B. *et al.* Ultralow-threshold electrically pumped quantum-dot photonic-crystal nanocavity laser. *Nat. Photonics* **5,** 297–300 (2011).

**Acknowledgements**

This research was supported by the Basic Science Research Program through the National Research Foundation of Korea (NRF) funded by the Ministry of Science, ICT & Future Planning (2015R1C1A1A01053117). This work was also supported by the Pioneer Research Center Program through the National Research Foundation of Korea (NRF) funded by the Ministry of Science, ICT & Future Planning (2014M3C1A3052580). This work was supported by National Research Foundation of Singapore (NRF-CRP12-2013-04 and Singapore MIT Alliance for Research and Technology's Low Energy Electronic Systems (LEES) IRG), Innovation Grant from SMART's Innovation Centre and SMA3 Fellowship. The simulation work was supported by US AFSOR Grant No. Grant: FA9550-15-1-0388. The authors thank D. Sukhdeo and R. Made for the fruitful discussions.

**Author contributions**

D.N. and K.S. conceived the initial idea of the project. S.B., K.H. and E.F. prepared GOI substrates. H.Q. and W.H. carried out electron-beam lithography. S.B. fabricated the samples and performed the Raman measurements. D.K. and C.O. carried out the optical measurements. Under the guidance of C.T. and D.N., S.B., D.K. and C.O. performed data analysis. S.G. and K.S performed the bandstructure simulations and gain modeling. Y.K., D.M. and Y.J. designed the laser structure by performing numerical analysis, finite-element method (FEM) and FDTD simulations, respectively. S.B., D.K., C.O., S.G., C.T. and D.N wrote the draft manuscript. C.T. supervised the entire part of material growth and sample fabrication. D.N. supervised the entire project.

Figures

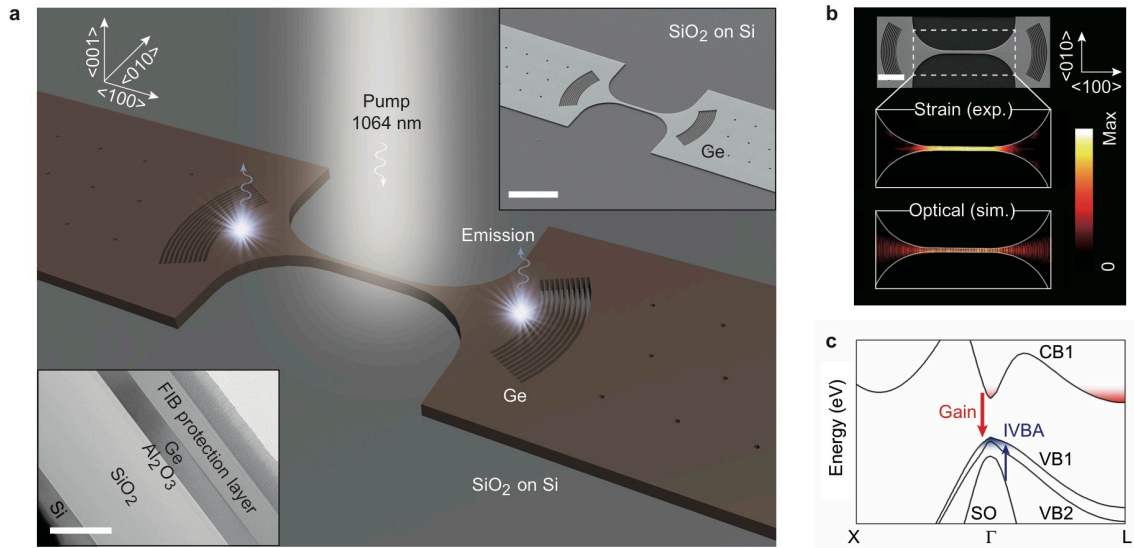

**Figure 1 | Design of strained Ge nanowire lasers. a,** Schematic illustration of a typical Ge nanowire laser consisting of a strained nanowire surrounded by a pair of DBRs on the stressing pads. The strained nanowire along the <100> direction is photo-excited with a 1064-nm pulsed laser, and the stimulated emission is collected at a DBR. Top inset: corresponding SEM image. Scale bar, 10 μm. Bottom inset: cross-sectional TEM image of the GOI structure. Scale bar, 0.5 μm. **b,** Top: top-view SEM image. Scale bar, 5 μm. Middle: 2D strain map measured by Raman spectroscopy showing a highly uniform strain distribution over the entire nanowire gain medium. Bottom: 2D optical field distribution calculated by FDTD simulation. A strong spatial overlap between strain and optical fields is achieved in our unique design. **c,** Calculated bandstructure of 1.6% uniaxial strained Ge. Two major optical processes, gain and IVBA, are clearly labeled.

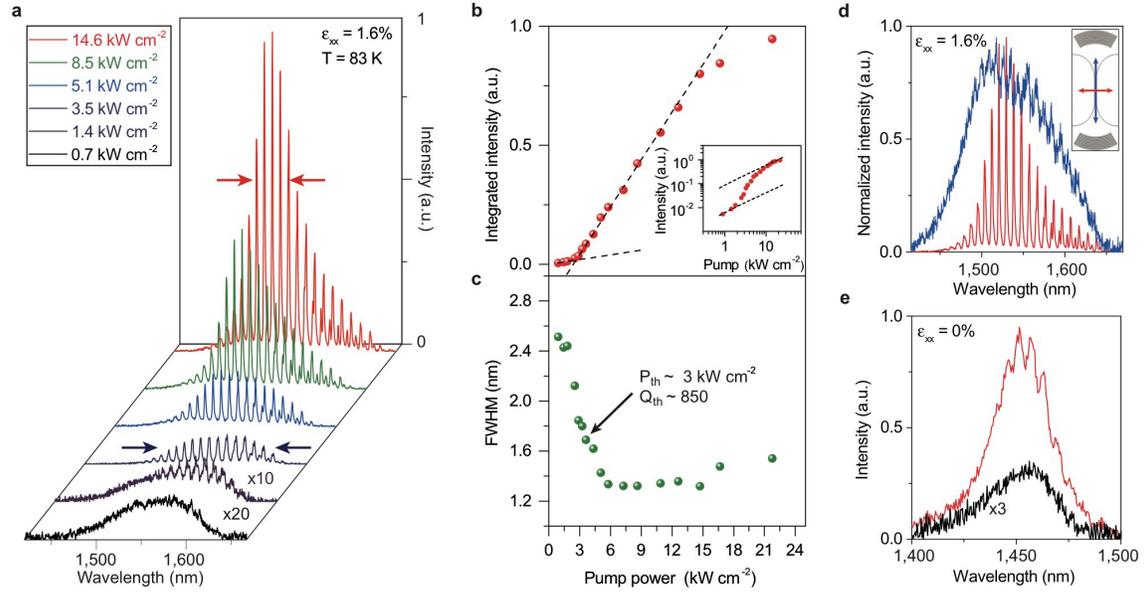

**Figure 2 | Lasing characteristics from strained Ge nanowires at 83 K. a,** Power-dependent photoluminescence spectra of a 1.6%-strained Ge nanowire with DBRs showing a gradual transition from broad spontaneous emission to multimode lasing oscillation (threshold, 3.0 kW cm$^{-2}$). The first and second spectra are multiplied by a factor of 20 and 10, respectively, for clarity. The arrows indicate an emission bandwidth of ~150 nm near the threshold (blue) and of <50 nm in the lasing regime (red). **b,** Integrated photoluminescence intensity versus optical pump power. The black dashed lines represent the linear fit to the experimental data indicating a clear threshold knee behavior. Inset: corresponding double-logarithmic plot showing nonlinear response to pump power represented by an S-shaped curve. **c,** The linewidth evolution of the lasing mode at 1530 nm as a function of pump power. The linewidth narrows from ~2.5 nm to ~1.3 nm. **d**, Normalized polarization-dependent spectra collected at 14.6 kW cm$^{-2}$, showing a highly anisotropic gain property of strained Ge nanowires. The emission polarized parallel to the strain axis (blue) does not show the optical amplification. **e,** Photoluminescence spectra of the unstrained structure taken at 0.7 kW cm$^{-2}$ (black) and
17

6.3 kW cm$^{-2}$ (red) pump powers, showing no lasing action. The spectrum for the pump power of 0.7 kW cm$^{-2}$ is multiplied by a factor of 3.

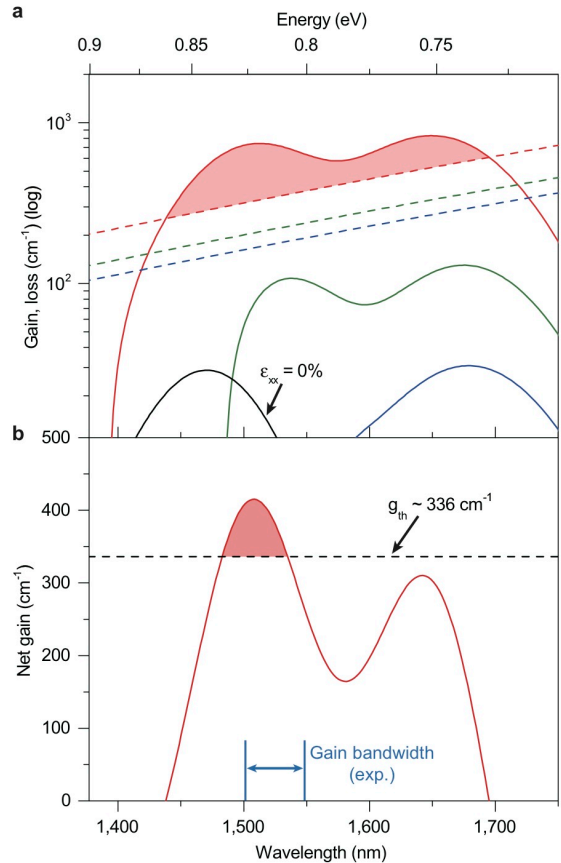

**Figure 3 | Theoretical modeling for gain and loss in strained Ge at 83 K. a**, Calculated gain (solid line) and loss (dashed line) for 1.6% strained Ge at injection densities of $4 \times 10^{19}$ cm$^{-3}$ (blue), $5 \times 10^{19}$ cm$^{-3}$ (green), and $8 \times 10^{19}$ cm$^{-3}$ (red). The peak optical net gain is ~ 415 cm$^{-1}$ at ~1510 nm. The gain for unstrained Ge at an injection density of $8 \times 10^{19}$ cm$^{-3}$ (black solid line) is overwhelmed by the loss. **b,** Calculated net gain spectrum for an injection density of $8 \times 10^{19}$ cm$^{-3}$. The gain band centered at ~1510 nm only surmounts the experimental threshold gain of 336 cm$^{-1}$. The measured gain bandwidth is also presented as a blue arrow.

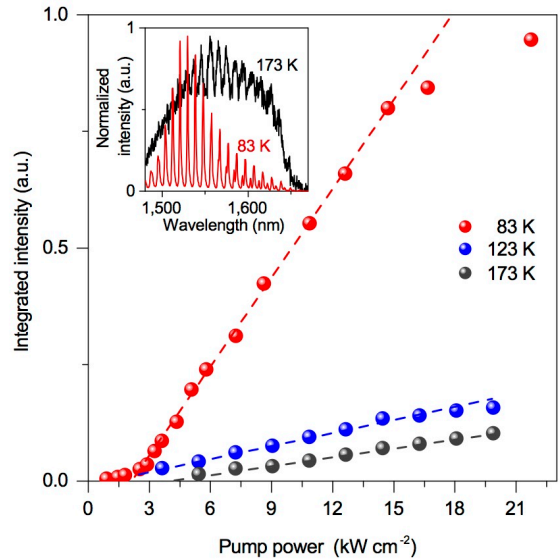

**Figure 4 | Temperature-dependent emission characteristics.** Integrated output intensity versus pump power of a 1.6% strained structure at temperatures of 83 K (red), 123 K (blue), and 173 K (black). While the dataset for 83 K manifests a nonlinear lasing behavior, no superlinear output increase is clearly observed for 123 K and 173 K. Inset: The normalized emission spectra collected at 83 K (red) and 173 K (black) under the pump power of 14.6 kW cm$^{-2}$. In contrast to the dominant lasing modes for 83 K, the observed cavity modes for 173 K are highly suppressed due to the absence of optical amplification.

# Supplementary Information

## Low-threshold optically pumped lasing in highly strained Ge nanowires

Shuyu Bao[1,2†], Daeik Kim[3†], Chibuzo Onwukaeme[3†], Shashank Gupta[4†], Krishna Saraswat[4], Kwang Hong Lee[2], Yeji Kim[3], Dabin Min[3], Yongduck Jung[3], Haodong Qiu[1], Hong Wang[1], Eugene A. Fitzgerald[2], Chuan Seng Tan[1,2*] and Donguk Nam[1,3*]

[1]School of Electrical and Electronic Engineering, Nanyang Technological University, 50 Nanyang Avenue, Singapore 639798
[2] Singapore-MIT Alliance for Research and Technology (SMART), 1 CREATE Way, #09-01/02 CREATE Tower, Singapore 138602
[3]Department of Electronic Engineering, Inha University, Incheon 402-751, South Korea
[4]Department of Electrical Engineering, Stanford University, Stanford, California 94305, USA

[†]These authors contributed equally to this work.
[*]E-mail: tancs@ntu.edu.sg; dnam@ntu.edu.sg

1. Stiction for heat conduction and finite-element method (FEM) modeling

2. Geometrically tunable strain and FEM modeling

3. Germanium-on-insulator (GOI) and laser structure fabrications

4. Design of optical cavity and FDTD simulations

5. Gain and loss modeling

6. References

# 1. Stiction for heat conduction and finite-element method (FEM) modeling

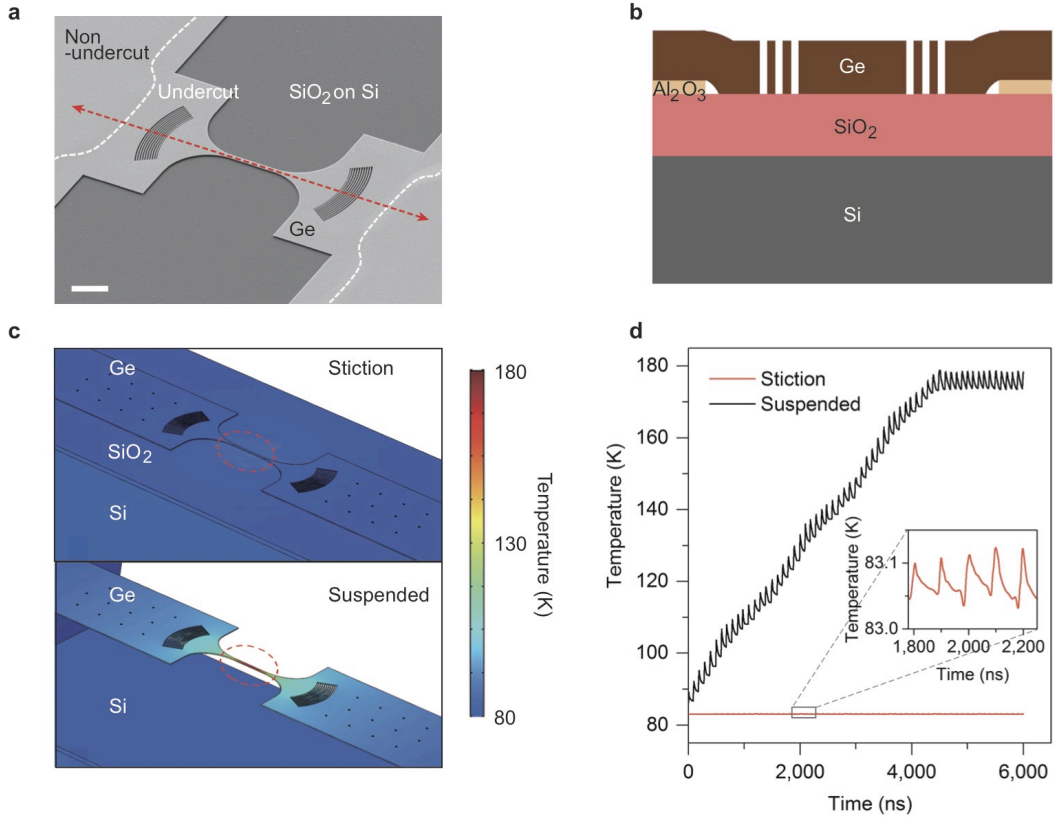

**Figure S1. a,** Scanning electron microscopy (SEM) image of a typical laser structure highlighting the undercut region. Scale bar, 5 μm. **b,** Cross-sectional schematic along the red dashed line of Fig. S1a. The germanium (Ge) nanowire gain medium is in close contact with the underlying silicon dioxide (SiO$_2$). **c,** Stationary FEM thermal simulation for temperature distribution in the stiction and suspended structures under a pulsed optical pumping of ~7 kW cm$^{-2}$ with a spot size of 15 μm. **d,** Time-dependent FEM thermal simulation for temperature variation as a function of time for the stiction and suspended structures under the same condition as Fig. S1c.

In our design, we intentionally bring the germanium (Ge) layer into contact with the underlying silicon dioxide (SiO$_2$) layer during the fabrication process (Supplementary Section 3). While the SiO$_2$ layer can effectively confine the optical mode within the Ge layer owing to a large refractive index difference between Ge and SiO$_2$, the heat accumulation by optical pumping can be significantly minimized in our architecture since the SiO$_2$ layer provides additional heat conducting paths towards the thick silicon (Si) substrate.

An SEM image of a finalized Ge laser structure in Fig. S1a clearly shows the boundary between the non-undercut and undercut regions. The undercut region is stuck onto the underlying SiO$_2$ layer, which allows superior heat conduction compared to the air gap employed in most other strained Ge resonator structures[1–3]. Figure S1b presents a cross-sectional schematic along the red dashed line in Fig. S1a to highlight the stiction of the top Ge layer.

FEM simulation is performed to investigate the heat conduction in our stiction structure and the conventional suspended structure. The center of nanowire is illuminated with a pulsed optical pumping of ~ 7 kW cm$^{-2}$ with a spot size of 15 μm which is above the pump power at the lasing threshold in our experiments (~3 kW cm$^{-2}$). The pulse duration and repetition period are 20 ns and 100 ns, respectively. The simulation temperature is set to 83 K to mimic the condition of our experiments.

Figures S1c and d show stationary and time-dependent FEM simulations showing temperature distributions in the two structures under the aforementioned condition. While the temperature of the suspended structure is quickly elevated up to 180 K owing to optical pumping, our stiction structure shows almost negligible temperature increase (only ~0.1 K) at the same condition, thus providing the evidence of an excellent thermal conduction in our architecture. Therefore, our unique design employing stiction plays a pivotal role for the achievement of low-threshold lasing from strained Ge by allowing superior thermal management in our structure.

\

## 2. Geometrically tunable strain and FEM modeling

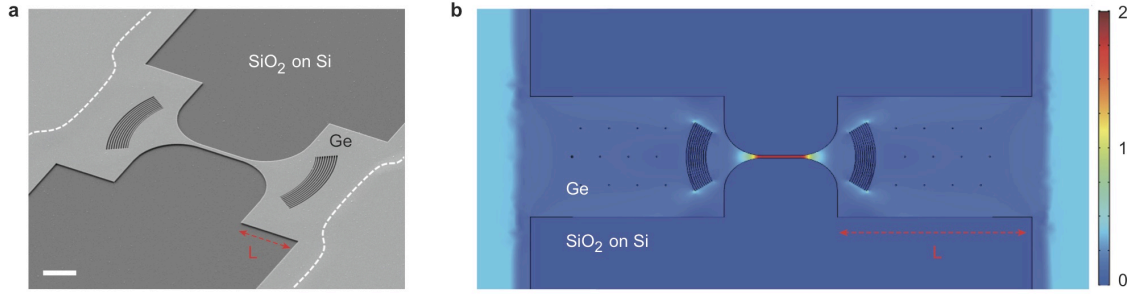

**Figure S2. a,** SEM image highlighting the length of the undercut region, L, for tuning the level of strain in Ge nanowires. Scale bar, 5 μm. **b,** FEM mechanical simulation showing a highly uniform strain distribution in the Ge nanowire.

In our design, we employ the strain amplification technique first introduced in ref. 4, and integrate an optical cavity in the stressing pads to provide optical feedback to the gain medium. Figure S2a shows an SEM image of an entire geometry highlighting the undercut length, L, which is used to adjust the level of strain by a conventional lithography. It is worth mentioning that by having multiple structures with different L on a single chip, one can achieve a number of lasers operating at different spectral ranges without complicated material growth such as for compound semiconductor lasers for which the composition of materials determines the operating wavelengths of lasers. Figure S2b presents a simulated 2D strain distribution by FEM mechanical simulation. In the simulation, a pre-existing film strain is set to be 0.2%. As a result of strain redistribution in the undercut structure, an amplified strain of ~2% exists within the entire gain medium uniformly. This allows us to achieve an extremely homogeneous gain medium which also plays an important role in enabling the lasing action in our highly strained Ge lasers.

## 3. Germanium-on-insulator (GOI) and laser structure fabrications

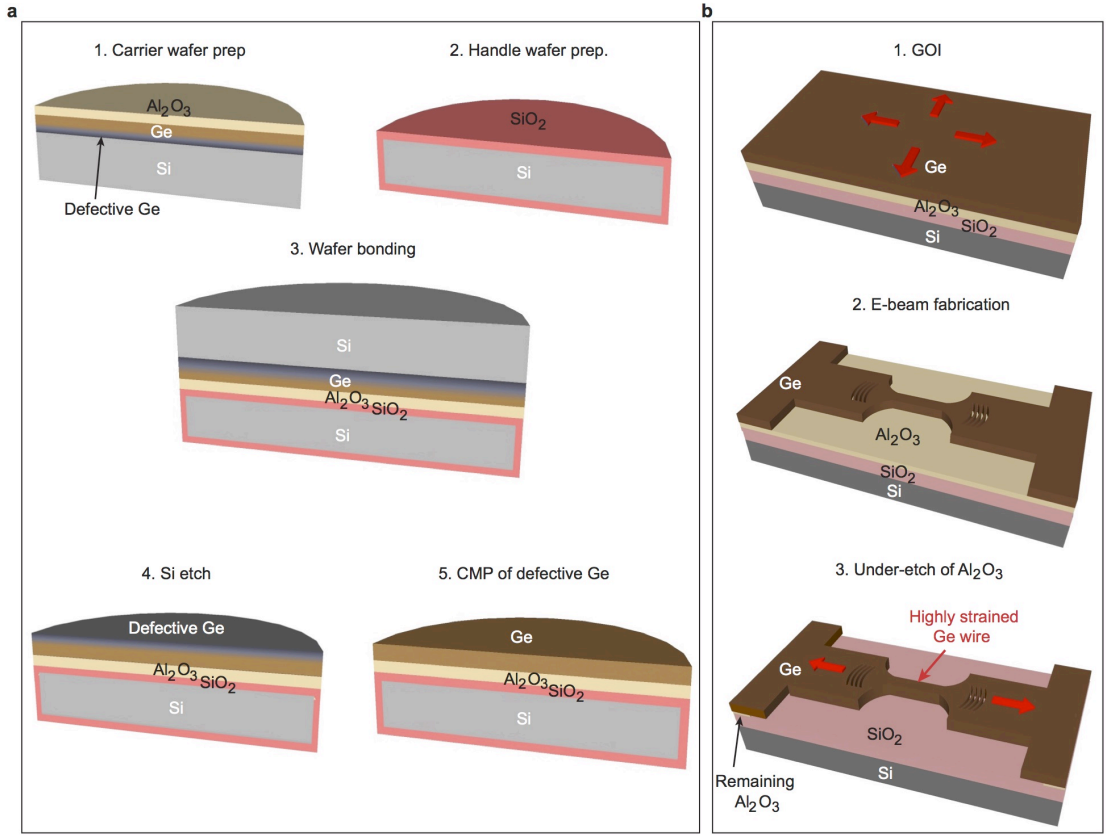

**Figure S3. a,** Schematics of the GOI substrate fabrication process. **b,** Schematics of the laser structure fabrication process.

Our GOI substrate for the Ge nanowire laser was made via epitaxy and wafer bonding (Fig. S3a), and the laser structure was then fabricated on GOI by lithography and etching processes (Fig. S3b). A bulk Ge layer is directly grown on an 8-inch Si (100) substrate using Aixtron MOCVD reactor. A low-temperature/high-temperature (LT/HT) two-step growth mode is used to grow a 50-nm LT Ge seed layer at 400 ˚C with heavy arsenic (As) doping and a 2.2-μm HT Ge layer at 650 ˚C[5]. During the HT growth, phosphorus (P) doping is introduced to the top 900-nm Ge layer to enable a uniform doping profile with a target P-doping concentration of ~ $6 \times 10^{18}$ cm$^{-3}$. Additionally, thermal cycling is applied to improve the epi-layer quality, leading to a threading dislocation density (TDD) of ~ $4.6 \times 10^6$ cm$^{-2}$. After removing the top 100-nm Ge layer by chemical mechanical polishing (CMP) step for a smooth surface for bonding, a 50-nm Al$_2$O$_3$ sacrificial layer is deposited on the Ge surface by atomic layer deposition (ALD). Then the Ge-on-Si wafer is directly bonded at room temperature to an 8-inch Si (100) handle wafer with a 1-μm thick thermal oxide ($SiO_2$) layer, followed by a post-bonding annealing at 300 ˚C for 3 hours to enhance the bonding strength. After removing the carrier Si by grinding and selective chemical etching in Tetramethylammonium hydroxide (TMAH), the Ge layer is transferred to the handle wafer, which forms a GOI substrate. Then the Ge layer of the GOI is thinned down by CMP to the desired thickness of 220 nm with a surface roughness of < 0.2 nm. The 8-inch GOI wafer is diced into 1 × 1 $cm^2$ pieces for device fabrication. The Ge nanowire laser structure is defined by the e-beam lithography (EBL), and its pattern is transferred to the Ge layer by reactive etching (RIE) using $Cl_2$ gas. The dry etch stops at the $Al_2O_3$ layer, and the sample is wet etched in 30%KOH solvent to selectively remove the sacrificial $Al_2O_3$ layer forming the undercut structure. The releasing process causes the strain redistribution and amplifies the tensile strain in the nanowire[4,6,7]. Then the sample is dimmed in 100% IPA. We finalize the laser fabrication by contact drying on a hotplate, which allows the nanowire to be in contact with the $SiO_2$ layer rather than suspended in air, resulting in a better heat conduction in our device.

## 4. Design of optical cavity and FDTD simulations

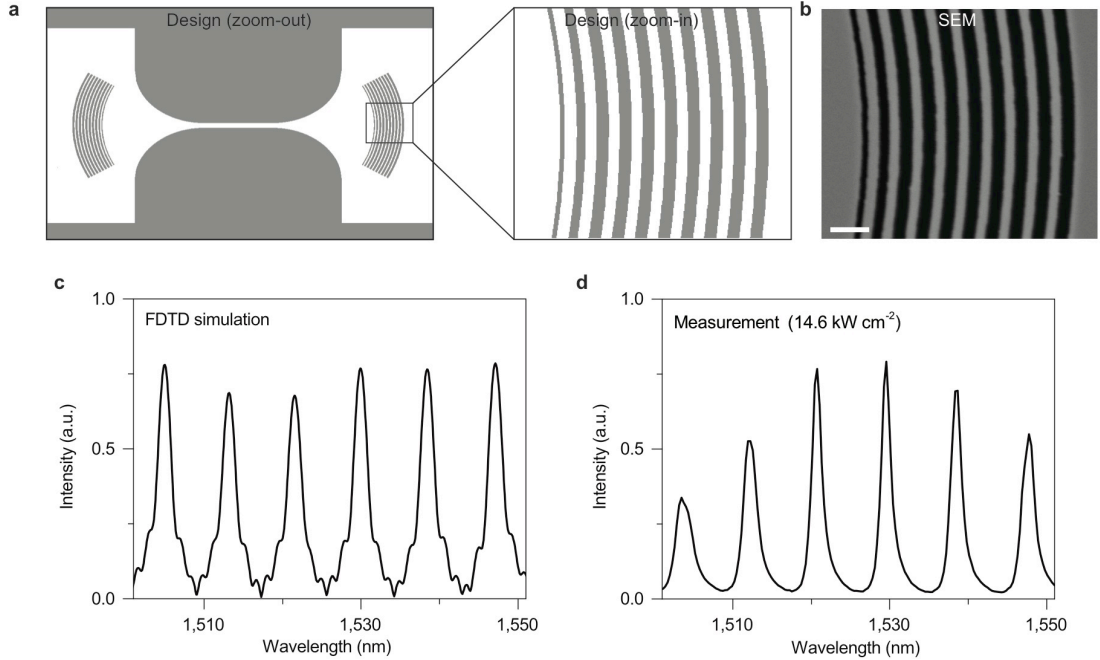

**Figure S4. a,** Design of our laser structure. A zoomed-in image of the distributed Bragg reflector (DBR) is also displayed. **b,** Zoomed-in SEM image of the DBR section of a fabricated device at the identical scale to the design image of Fig. S4a. Scale bar, 1 μm. **c,** Simulated spectrum of our structure (Fig. S4a) using finite-difference time-domain (FDTD) simulation. **d,** Measured spectrum at an optical pumping density of 14.6 kW cm$^{-2}$.

The laser structure consists of an 8-μm long, 700-nm wide Ge nanowire surrounded by two stressing pads containing a pair of distributed Bragg reflector (DBR) mirrors (Fig. S4a). The DBR is carefully designed to maintain the homogeneity in the strain distribution along the nanowire. The DBR contains 10 air trenches with width and period of 189 nm and 379 nm, respectively. The width of the first air trench (narrowed to 65 nm) and the curvature of circular arcs are optimized to obtain a high optical Q factor. Figure S4b displays an SEM image of the DBR section of a fabricated device showing an almost identical image to the original design.

Figure S4c presents a simulated spectrum of cavity modes in 1500-1550 nm with a mode spacing of ~8 nm which is in a good agreement with our experimental result (Fig. S4d). While the simulated Q factors are well above 1,000, the experimental Q factor at the threshold is measured to be ~850 and this discrepancy may be attributed to the sidewall roughness of our DBRs introduced during fabrication processes[8]. We believe that improving the cavity design and fabrication processes will allow us to further reduce the lasing threshold.

**5. Gain and loss modeling**

*a. Detailed description on gain and loss modeling*

We perform theoretical modeling to obtain the gain spectrum of uniaxial tensile strained Ge. Empirical pseudopotential method (EPM) is used to compute the bandstructure of strained Ge. EPM is an attractive approach because it allows for the computation of bandstructures with relatively small number of empirical parameters. In calculation for bandstructure, the single electron Hamiltonian is expressed as[9]:

$$H(G,G') = -\frac{h^2}{2m}\nabla^2 + V_{loc}(|G-G'|) + V_{nloc}(G,G') + V_{so}(G,G'), \quad (1)$$

where $V_{loc}$, $V_{nloc}$, and $V_{so}$ represent the local, nonlocal, and spin orbit contributions to the pseudopotential, respectively.

After calculating the bandstructures, the transition rates between different bands in Ge are calculated by using Fermi's golden rule. The absorption (or gain) due to band-to-band transitions is calculated using:

$$\alpha = C_0 \sum_k \delta(E_1(k) - E_2(k) - \hbar\omega)(f_1 - f_2)|p_{12}\cdot\hat{e}|^2, \quad (2)$$

where $\delta(E_1(k)-E_2(k)-\hbar\omega)$ computes the joint density of states (JDOS) between bands 1 and 2 corresponding to transitions energy $\hbar\omega$. $(f_1-f_2)$ represents the Fermi inversion factor where $f_1$ and $f_2$ are Fermi functions. The quantity $|p_{12}\cdot\hat{e}|^2$ is the squared momentum matrix element for a particular wave vector k and $\hat{e}$ is the polarization. $C_0=\pi e^2/(m^2\varepsilon_0 cn)$ is a constant, where $e$ is the electron charge, $\varepsilon_0$ is the vacuum permittivity, $c$ is the light speed in vacuum, and $n$ is the refractive index. The three quantities – JDOS, Fermi inversion factor and momentum matrix elements are calculated at each point in the first Brillouin zone (FBZ) and their product is then summed throughout the FBZ to obtain the material absorption (or the material gain) spectrum. For all gain curves in the main article and Supplementary Information, we only consider one polarization along the perpendicular direction to the nanowire axis since we observe the optical amplification only for such a polarization direction.

To achieve the optical net gain in Ge, the material gain should overcome the combined material loss consisting of free electron absorption (FEA) and inter-valence band absorption (IVBA)[6]. For the losses at room temperature, we take the same formalism used in Ref. 6, but divide the total loss by a factor of 3 assuming isotropic losses since we only consider one polarization direction for the material gain. It is highly required to investigate the polarization dependence of material losses to more precisely understand the gain and loss behavior in Ge. Since IVBA is strongly dependent on temperature[10], we take an experimentally measured IVBA at 95 K which is close to the temperature at which we observed lasing. By fitting IVBA data of Ge with p-type doping concentration of 1 × 10$^{19}$ cm$^{-3}$ at 95 K[10], we obtain:

$$\alpha_{IVBA}(\hbar\omega) = 1.13 \times 10^{-14} N_{h,tot} e^{-8.786(\hbar\omega)}, \tag{3}$$

where $\hbar\omega$ is the photon energy, $N_{h,tot}$ is the total hole density. $\alpha_{IVBA}$ is in units of cm$^{-1}$, $\hbar\omega$ is in units of eV, and $N_{h,tot}$ is in units of cm$^{-3}$. The curve for 95 K is more strongly dependent on temperature than the counterpart for 300 K, and this can be attributed to the difference in Fermi-Dirac functions at two different temperatures. We use this experimental IVBA for our modeling at 83 K, and for FEA, we employ the same formalism used for 300 K[6] since FEA is not a strong function of temperature.

*b. Modeling results*

Figure S5a shows the separate contributions of FEA and IVBA to the combined loss for a carrier injection density of 7 × 10$^{19}$ cm$^{-3}$. At 7 x 10$^{19}$ cm$^{-3}$ carrier density and at the experimental peak gain wavelength of ~1530 nm, IVBA is >214 cm$^{-1}$ whereas FEA is only <96 cm$^{-1}$. Therefore, IVBA contributes to the major part of the material loss and it is critical to minimize the IVBA by lowering operating temperatures to achieve optical net gain.

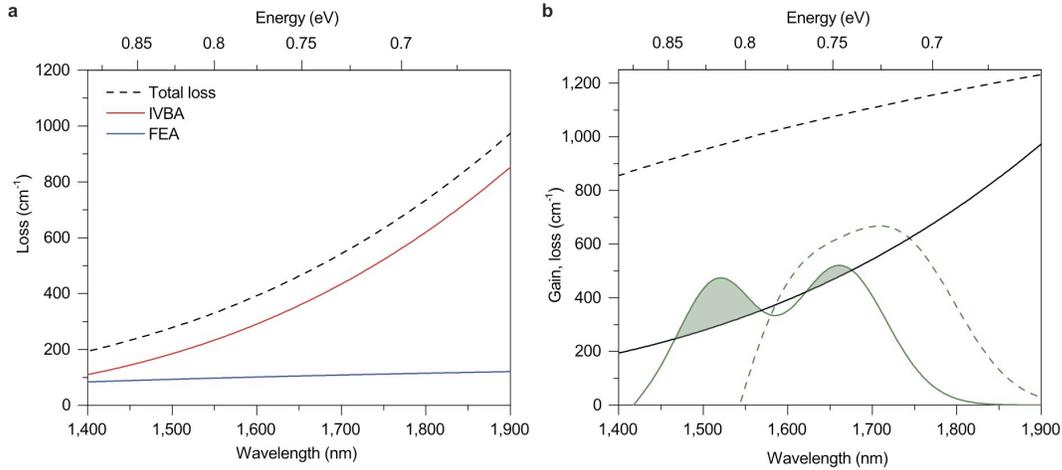

**Figure S5. a,** FEA and IVBA contributions to the total loss for a carrier injection density of 7 × $10^{19}$ cm$^{-3}$. **b,** Calculated gain (green) and loss (black) at 83 K (solid) and 300 K (dashed) for 1.6% strained Ge at a carrier injection density of 7 × $10^{19}$ cm$^{-3}$. The green shaded area corresponds to the positive net gain region.

Figure S5b shows the calculated gain (green lines) and loss (black lines) at two different temperatures of 83 K (solid lines) and 300 K (dashed lines) for 1.6% strained Ge. The carrier injection density and the doping density are 7 × $10^{19}$ cm$^{-3}$ and 6 × $10^{18}$ cm$^{-3}$, respectively. While the magnitudes of the peak material gain at two temperatures do not show a significant difference, IVBA can be greatly reduced at 83 K, thereby allowing the achievement of a net gain of ~180 cm$^{-1}$.

# 6. References

1.  El Kurdi, M. *et al.* Tensile-strained germanium microdisks with circular Bragg reflectors. *Appl. Phys. Lett.* **108,** 91103 (2016).

2.  El Kurdi, M. *et al.* Direct band gap germanium microdisks obtained with silicon nitride stressor layers. *ACS Photonics* **3,** 443–448 (2016).

3.  Geiger, R. Direct band gap germanium for Si-compatible lasing. PhD Dissertation, ETH Zürich (2016).

4.  Minamisawa, R. A. *et al.* Top-down fabricated silicon nanowires under tensile elastic strain up to 4.5%. *Nat. Commun.* **3,** 1096 (2012).

5.  Lee, K. H. *et al.* Reduction of threading dislocation density in Ge/Si using a heavily As-doped Ge seed layer. *AIP Adv.* **6,** 025028 (2016).

6.  Süess, M. J. *et al.* Analysis of enhanced light emission from highly strained germanium microbridges. *Nat. Photonics* **7,** 466–472 (2013).

7.  Nam, D. *et al.* Strain-induced pseudoheterostructure nanowires confining carriers at room temperature with nanoscale-tunable band profiles. *Nano Lett.* **13,** 3118–3123 (2013).

8.  Ellis, B. *et al.* Ultralow-threshold electrically pumped quantum-dot photonic-crystal nanocavity laser. *Nat. Photonics* **5,** 297–300 (2011).

9.  Gupta, S., Magyari-Köpe, B., Nishi, Y. & Saraswat, K. C. Achieving direct band gap in germanium through integration of Sn alloying and external strain. *J. Appl. Phys.* **113,** 73707 (2013).

10. Morozov, I. & Ukhanov, I. Effect of doping on the absorption spectra of p-Ge. *Sov. Phys. J.* **13,** 744–747 (1970).